\documentclass[]{raa_twocolumn}        

\usepackage{graphicx,times}             
\usepackage{natbib}
\usepackage{float}
\usepackage{soul}
\usepackage{subcaption}

\DeclareCaptionLabelFormat{figandsubfig}{Fig.~#2} 
\makeatletter
\renewcommand\p@subfigure{}
\makeatother
\usepackage{amssymb,amsmath}
\bibpunct{(}{)}{;}{a}{}{,}

\newcommand\pyIGIMF{{\tt pyIGIMF}}

\newcommand{\diff}[0]{\mathrm{d}}

\usepackage[pagebackref=true]{hyperref}

\begin{document}

  \title{The Initial Mass Function as the Equilibrium State of a Variational Process: why the IMF cannot be sampled stochastically
}

   \volnopage{Vol.0 (20xx) No.0, 000--000}      
   \setcounter{page}{1}          

   \author{Eda Gjergo$^{\star}$ 
      \inst{1,2}
   \and Zhiyu Zhang$^{\dagger}$ 
      \inst{1,2}
   \and Pavel Kroupa$^{\ddagger}$
      \inst{3,4}
   }

   \institute{School of Astronomy and Space Science, Nanjing University, 
   Nanjing 210093, China; {\it \\ $^{\star}$eda.gjergo@gmail.com, $^{\dagger}$zzhang@nju.edu.cn, $^{\ddagger}$pkroupa@uni-bonn.de
   }\\
        \and
             Key Laboratory of Modern Astronomy and Astrophysics (Nanjing University),\\
Ministry of Education, Nanjing 210093, People's Republic of China.\\
        \and
             Helmholtz-Institut f\"ur Strahlen und Kernphysik, Universit\"{a}t Bonn,\\
Nussallee 14-16, Bonn, North Rhine-Westphalia, D-53115, Germany.\\
        \and 
            Charles University in Prague, Faculty of Mathematics and Physics, Astronomical Institute,\\ V Hole\v{s}ovi\v{c}kách 2, CZ-180 00 Praha 8, Czech Republic.
\vs\no
   {\small Received 20xx month day; accepted 20xx month day}}

\abstract{ The stellar initial mass function (sIMF) is often treated as a stochastic probability distribution, yet such an interpretation implies Poisson noise that is inconsistent with growing observational evidence. In particular, the observed relation between the mass of the most massive star formed in an embedded cluster and the cluster's total stellar mass supports a deterministic sampling process, known as optimal sampling. However, the physical origin of optimal sampling has not been formally established in the literature.
In this work, we show that the stellar mass distribution implied by optimal sampling emerges from applying the Maximum Entropy principle to the fragmentation of star-forming clumps, whose structure is set by density-dependent cooling in the optically thin regime.
Here, the maximum entropy leads to unbiased distributions. 
By applying calculus of variations to minimize the entropy functional obtained assuming fragmentation, we recover the power-law form of the sIMF, and we show that any distribution deviating from the sIMF violates the Maximum Entropy principle. This work provides a first-principles foundation for the deterministic nature of star formation. Thus, the sIMF 
is the distribution resulting from a
maximally unbiased system. 
}

   \authorrunning{E. Gjergo, Z. Zhang \& P. Kroupa }            
   \titlerunning{Optimal Sampling from Maximum Entropy}  

   \maketitle

%
%

\section{Introduction}           
\label{sec:intro}

One of the most important distribution functions in astrophysics is the \emph{stellar initial mass function} (sIMF\footnote{This is most commonly known as the IMF, but here we draw the distinction between the sIMF of a single stellar population and the galaxy-wide IMF which involves a whole galaxy \citep[see][for further details]{kroupa+2024}.}), which refers to the number distribution of stellar masses generated during a single star~formation episode.
The sIMF is described by the following distribution:
\begin{equation}\label{eq:IMFdistribution}
    \xi(m) = \frac{\diff N}{\diff m} \, ,
\end{equation}
where $N$ refers to the total number of stars generated, and $m$ to their mass. The typical sIMF pattern in Milky-Way-like galaxies can be expressed as a broken~power-law (see e.g. \citealt{Kroupa2001}, hereinafter K01, also known as the canonical IMF, see also \citealt{Jerabkova+2025}):
\begin{equation}\label{eq:canonicalIMF}
    \xi_{\rm K01} (m) = k_{\star,j} \, m^{-\alpha_{\rm j}}  \, ,
\end{equation}
where the slopes, $\alpha_j$, are obtained empirically according to the mass interval. In K01, which analyzed and formalized the canonical values for the Milky Way, $\alpha_{m < 0.5 \, {\rm M}_{\odot}} = 1.3$ and $\alpha_{m > 0.5 \, {\rm M}_{\odot}} = 2.3 \,$. 
Molecular clumps are here defined as the densest self-gravitating substructures of molecular clouds that collapse to form stars \citep{McKeeOstriker2007, BerginTafalla2007}. All stars formed within a clump constitute a single stellar population (SSP) produced in one star formation episode \citep[e.g.,][]{Lada2010}. 
We do not address the core mass function because it is a transient distribution: at any given epoch it describes only the subset of cores present at that stage and is not in one-to-one correspondence with the final sIMF \citep[e.g.,][]{Zhou+2025}. The sIMF, on the other hand, represents the complete stellar mass distribution arising as a SSP.
Within a few Myr, the embedded cluster disperses and eventually dissolves through dynamical evolution.
\citep[see ][for a review]{kroupa+2024}.

The sIMF is sometimes interpreted as a stochastic probability density distribution 
\citep[e.g.][and references therein]{Elmegreen1997, Bastian2010}. 
If sampling were stochastic, the sIMF would represent the mean of several hypothetical realizations of single stellar populations (SSP). Any given SSP would not match the sIMF exactly.
Moreover, Poisson sampling noise would be largest where counts are small at the high-mass end and smallest where counts are large at low masses. Consequently, low-count bins at high masses may occasionally be empty. 

However, observational data disfavor the stochastic interpretation (as further discussed in Sec.~\ref{sec:discussion}). 
Two lines of argument suggest this to be the case: (i) the relation between the embedded cluster stellar mass in stars, $M_{\rm ecl}$, and its most massive star, $m_{\rm max}$, shows little scatter across heterogeneous datasets \citep{weidner13, Weidner+2014, yan+23}; (ii) the high-mass sIMF slope distribution has a narrow peak at the Salpeter value, $\alpha_{\rm S55} \approx 2.35$ \citep[][]{kroupa2002}. 

The sIMF appears instead to be highly self-regulated, and connected to the macroscopic properties of a molecular clump (e.g., total gas mass, metallicity, etc.). 
If two clumps have the same physical properties, such as density and metallicity, in fact, they will produce identical stellar populations \citep{kroupa+2013, kroupa+2024}.

To differentiate between ``stochastic'' and ``deterministic'' scenarios of star formation, the strongest and unavoidable observational evidence comes from the  intrinsic scatter in the relation between $m_{\rm max}$, and $M_{\rm ecl}$   \citep[the $m_{\rm max}-M_{\rm ecl}$~relation is investigated in the following works:][]{weidner+2006, weidner+2010, weidner13, Weidner+2014, KirkMyers2011, Stephens+2017}. The intrinsic scatter of this relation is surprisingly small \citep{yan+23}.
An almost identical relation has been observed in statistical studies of dense cores in Galactic star-forming regions \citep[e.g.,][]{Xu2024}. This suggests that the mechanisms that regulate star formation, and shape the resulting sIMF, may originate as early as the formation of prestellar cores.

A deterministic scheme called \emph{optimal sampling} was introduced by \citet{kroupa+2013}. 
Optimal sampling is subject to two constraints:
\begin{align}
    \int_{m_{\rm min}}^{m_{\rm max}} m \, \xi (m) \, \diff m & = M_{\rm ecl} \, ,\label{eq:opt_mass}\\
    \int_{m_{\rm max}}^{m_{\rm max *}} \xi(m) \, \diff m & = 1 \, , \label{eq:opt_number}
\end{align}
where $M_{\rm ecl}$ is the embedded cluster total stellar mass, $m_{\rm max}$ is the mass of the most~massive~star in the embedded cluster, and $m_{\rm min}$ is the mass of the least~massive star. 
The upper bound $m_{\rm max*} \approx 150\,{\rm M}_{\odot}$ is the physically accessible maximum boundary for the highest stellar mass that can form, as discussed in \citet{Gjergo+2025} (see also \citealt{kroupa+2013}, \citealt{kroupa+2024}). Eq.~\ref{eq:opt_mass} ensures that the integral of the mass-weighted sIMF yields the total cluster mass. 
Eq.~\ref{eq:opt_number} specifies that exactly one star is drawn from the interval $[m_{\rm max}, m_{\rm max *}]$, and hence defines the most massive star in the cluster. Eqs.~\ref{eq:opt_mass} and \ref{eq:opt_number} hold also for very massive clusters where $m_{\rm max} \rightarrow m_{\rm max *}$, but these equations are particularly relevant for low-mass clusters. In such low-mass systems, these constraints prevent low-mass, low-density molecular clumps from forming massive stars. They also ensure that stellar masses follow the sIMF closely, without any gaps or Poisson noise. 

The formation of massive stars provides a stringent test for distinguishing between stochastic and optimal sampling. 
If the sIMF were a probability distribution, it would be possible, although highly unlikely, that, e.g., a $1000 \, {\rm M}_{\odot}$~embedded cluster could generate  an 80 or even $100 \, {\rm M}_{\odot}$~star. With optimal sampling, such outcomes are forbidden. Here, the sIMF does not describe a probabilistic process, but instead prescribes how the physical state of the system partitions the available clump mass into stars. This interpretation precludes any stochastic over- or under-representation of high- or low-mass stars. 

Although optimal sampling is in excellent agreement with observations \citep{kroupa+2024}, a rigorous derivation from first principles has not yet been formally investigated in the literature,
beyond the suggestion that optimal sampling might be a consequence of self-regulation by feedback during the star formation process \citep{kroupa+2013, Vazquez-Semadeni+2024}.

In this work, we show that 
optimal sampling, in the form of Eqs.~\ref{eq:opt_mass}~and~\ref{eq:opt_number}, results from applying the Maximum Entropy principle to gas fragmentation. Here we define fragmentation as structure formation from density-dependent cooling. 
The Maximum Entropy principle (MaxEnt) is a rule to obtain the probability distribution of a system subject to macroscopic constraints when its micro-states are unknown. 
Under this principle, of all the distributions that satisfy the macroscopic constraints, the representative distribution is the one with the largest information entropy. 
MaxEnt was first formalized by \citet{Jaynes1957a, Jaynes1957b}, where the principle was applied to derive the Boltzmann distribution though information-entropy maximization rather than from dynamical arguments in statistical mechanics. The MaxEnt solution is the least biased because it introduces no structure beyond the imposed macroscopic constraints \citep[for a historical overview, see][]{Presse+2013}. Equivalently, it selects the minimum-information distribution consistent with those constraints.

The Boltzmann distribution, like other equilibrium distributions obtained in classical mechanics, describes a state of high information entropy\footnote{In this text, unless otherwise noted, entropy refers always to “information entropy”.} \citep{CarcassiAidala2020}. As such, it has no intrinsic stochasticity in the fundamental laws that govern the system. This contrasts with the Copenhagen interpretation of quantum mechanics, where measurement outcomes are intrinsically stochastic \citep{Bohr1928, vonNeumann1955}. At the same time, quantum systems have fewer available states and therefore lower information entropy \citep{CarcassiAidala2020}. Rather, the distribution is an emergent feature from the complexity of the ensemble of micro-states involved. Such distributions encode incomplete information about micro-states rather than intrinsic randomness. 
Maximizing entropy, then, means maximizing our lack of knowledge of the state of the system, and results in the least biased distribution \citep[e.g.,][]{Kesavan2009}.

In Sec.~\ref{sec:method} we show that, indeed, optimal sampling from \cite{kroupa+2013} is a corollary of a maximally entropic distribution resulting from the fragmentation of molecular clumps. We visualize one realization of the sIMF under the hypothesis it is a probability density function, and we compare to the sIMF for a similar embedded cluster, but computed under optimal sampling. In Sec.~\ref{sec:discussion} we present past attempts at interpreting the features of the sIMF, with motivations and shortcomings, and lastly we summarize our findings in Sec.~\ref{sec:conclusion}.

\section{Derivations and Analysis}
\label{sec:method}

We present the maximum entropy formalism in Sec.~\ref{sec:MaxEnt}. We discretize Eq.~\ref{eq:IMFdistribution} in Sec.~\ref{sec:SF}. 
In Sec.~\ref{sec:fragmentation} we build the most fundamental constraint on star formation, namely that stars form from the fragmentation of molecular clumps. In Sec.~\ref{sec:sIMF_MaxEnt} we apply the Maximum Entropy formalism to the distribution function of stellar masses of a single stellar population (i.e., the discrete sIMF), and we show that the corollary properties of the resulting distribution consist of the optimal sampling conditions from Eqs.~\ref{eq:opt_mass}~and~\ref{eq:opt_number}.

\subsection{Maximum Entropy Formalism}\label{sec:MaxEnt}

We follow the maximum entropy formalism in the formulation presented by \citet{Kesavan2009}.  Let us consider a variable $X$ whose distribution is described by $\bf p$. The ensemble $(X, {\bf p}) = ((x_1,p_1), ...,(x_{\rm n}, p_{\rm n}))$ describes the discrete multivariate distribution of all possible realizations of $X$, and their probability of occurrence. The laws governing the behavior of $(X, {\bf p})$ may be deterministic, but they are inaccessible to the observer. However, the quantity $\bf p$ is subject to the natural constraint:
\begin{equation} \label{eq:probability_axiom}
    \sum_{i=1}^n p_i=1 \, .
\end{equation}

In Information Theory, the measure of entropy (the observer's uncertainty), $S$, for any ensemble whose distribution, $\bf p$, abides by Eq.~\ref{eq:probability_axiom} was found by \citet{Shannon1948} to be:
\begin{equation}\label{eq:entropy}
    S = - \sum_{i=1}^n p_i \, \ln p_i \, .
\end{equation}
Interestingly, within Jaynes' maximum entropy formalism it was found that 
for any admissible set of distributions $\{p_i\}$ that satisfy Eq.~\ref{eq:probability_axiom} and any additional linear constraints, the entropy functional in Eq.~\ref{eq:entropy} has a maximum on that set. This means that $\{p_i\}$ represents the state of greatest uncertainty about the micro-states of the system \citep[][their pp.1780]{Kesavan2009}.

The distribution $p_i$ can be obtained by applying calculus of variation to the following Lagrangian, $\mathcal{L}$ \citep{Jaynes1957a, Jaynes1957b}:

\begin{equation}\label{eq:lagrangian}
    \mathcal{L} = S - (\lambda_0 -1)\left(\sum_{i=1}^n p_i -1\right) - \sum_{r=1}^{t} \lambda_r \left(\sum_{i=1}^n p_i g_{ri} - a_r\right) \, ,
\end{equation}
where $\lambda_0, ..., \lambda_{t}$ are Lagrange multipliers. 
The zeroth Lagrange multiplier, $\lambda_0$, corresponds to the constraint from Eq.~\ref{eq:probability_axiom}, while indices $r=1,...,t$ span through all constraints applicable to the ensemble $(X, {\bf p})$:

\begin{equation} \label{eq:Lm_constraints}
    \sum_{i=1}^n p_i \, g_{ri} = a_r \, ,
 \end{equation}
where $a_r$ refers to the values of constraints $r$, while $g_{ri}$ are their moment constraints.

The distribution, $p_i$, of the system is then obtained through calculus of variation by minimizing the Lagrangian, Eq.~\ref{eq:lagrangian}:
\begin{equation}
    \frac{\partial \mathcal{L}}{\partial p_i} = 0 \,.
\end{equation}

\subsection{The formation of a single stellar population}\label{sec:SF}

In the case of an ensemble of newly formed stars in a molecular clump whose micro-states are maximally unknown to the observer, the resulting number distribution of their masses is also described by $(M, {\bf p})= ((m_1,p_1),...,(m_n, p_n))$. 
Each microstate refers to individual stars. We label microstates by stellar mass $m_i$, and assign each microstate a probability $p_i$. 
Before finding the relation between $\xi(m)$ and $p_i$, let us first discretize the sIMF, $\xi(m)$.

\citet{kroupa+2013} proposed the following scheme, later refined slightly by \citet{schulz15} and \citet{yan+2017}: let $\rm i = \{1, ..., \rm N \}$ refer to the index\footnote{$N$ and $n$ are used interchangeably ($N \equiv n$) as they both refer to the number of stars in the astrophysical and information theory formalisms respectively.} of a mass-ordered ensemble of stars  (i.e., $m_{\rm i+1} > m_{\rm i}$) generated by a molecular clump so that $m_1 \equiv m_{\rm min}$ is the least massive star and $m_{\rm N} \equiv m_{\rm max}$ is the most massive. 
The sequence of stellar masses is determined recursively by the distribution of stellar mass and its normalization: 

\begin{align}
 m_{\rm i+1} &= \int^{m_{\rm i+1}}_{m_{\rm i}} m\,\xi(m) \, {\rm d}m \, , \label{eq:m_i} \\   
 1 &= \int^{m_{\rm i+1}}_{m_{\rm i}} \xi(m) \, {\rm d}m \, ,
\end{align}
as originally presented in \citet{kroupa+2013}, their Eq.~9. Note that their framework is defined in reverse-mass order (i.e., $m_{\rm i+1} < m_{\rm i}$, with integration limits accordingly reversed). The two equations, Eq.~\ref{eq:opt_mass} and Eq.~\ref{eq:opt_number}, solved simultaneously, return the most massive star that may form in an embedded cluster, as well as the normalization constant of the sIMF. Therefore, the optimal sampling condition says that for any given embedded cluster there must be one most massive star, and that this most massive star defines in turn the normalization constant of the sIMF.

This method defines the inverse-mass-ordered list of stellar masses, $m_i$. We now discretize the sIMF by defining:
\begin{equation}\label{eq:xi_i}
    \xi_i = \xi(m_i) \, .
\end{equation}
We may define $\xi_i$ in terms of a normalized discrete distribution $p_i$, such that:
\begin{align}\label{eq:p_i}
    p_i &\equiv \int_{m_i}^{m_{i+1}} p(m)\, \diff m 
    \approx p(m_i)\, \Delta m_i  \nonumber \\
    &= \frac{\xi(m_i)}{N}\, \Delta m_i 
    = \frac{k_{\star}}{N}\, m_i^{-\alpha} \Delta m_i \, .
\end{align}
Here $p(m)$ is the probability density in mass, $\xi(m) = k_{\star} m^{-\alpha}$ is the sIMF, and $k_{\star}$ is the usual normalization constant.
This relation holds in the continuum limit because the sum over all $p_i$ is 1, while the integral of $\xi(m)$ over the stellar mass range returns $N$, the total number of stars. $\Delta m_i$ is the mass interval associated with bin $i$, for example $\Delta m_i = m_{i+1}-m_i$ for the ordered mass sequence.

\subsection{Molecular Clump fragmentation}\label{sec:fragmentation}

The state of a molecular clump is defined by its physical properties, such as gas mass, $M_{\rm clump}$, metallicity ($Z$), temperature, cosmic-ray flux, specific angular momentum, magnetic field and other physical parameters \citep[][]{Elmegreen1989}.
As explained in \cite{kroupa+2024}, evidence suggests that $M_{\rm clump}$ is the dominant factor, and $Z$ may play a secondary role. The remaining parameters have negligible influence for typical clumps. 
We therefore neglect these secondary dependencies and treat the clump mass as the sole relevant variable. The relation between clump gas mass, $M_{\rm clump}$, and embedded cluster stellar mass, $M_{\rm ecl}$, is given by: 
\begin{equation}
 M_{\rm clump} = M_{\rm ecl}/\epsilon \, ,   
\end{equation}
where $\epsilon=0.3$ is the clump-scale star formation efficiency \citep[][and references therein]{zhou+2024}. For
simplicity, we assume $\epsilon$ to be constant, which, as has been recently shown, appears to be universal against various turbulent environments \citep{Jiao2025baobab}. The dependence on the neglected parameters will be investigated in future works.

A molecular clump must fragment in order to form stars  \citep[e.g.,][]{zinnecker1989, jappsen+2005, Goodwin+2008,Zhang2009, elmegreen2011b, anathpindika2013, beuther+2024}.
Its mass is progressively divided through hierarchical fragmentation: each collapsing structure splits into two or more substructures, collapses into filaments and fibers \citep{Andre+2014, hacar+2017}, which in turn fragment further until the thermal Jeans length and mass are reached, forming molecular cloud cores along the filaments \citep[e.g.,][]{Myers2009,Myers2011,Andre+2014,  Zhou+2025}, at which point they may form stars \citep{Zhang2009,Morii2024,Li2024}.  

In such a hierarchy, each daughter clump inherits a fraction of the parent mass, set by local density and cooling conditions, and the mass of a fragment after several levels is the product of the mass fractions along its branch rather than a sum of fixed mass increments. 
This process is therefore multiplicative rather than additive. For a hierarchical, scale invariant, multiplicative process, the geometric mean provides a natural characteristic mass scale \citep{Siegel1942}. Fixing the geometric mean encodes the scale-invariance and hierarchical nature of a system:
\begin{equation}\label{eq:geo_mass}
    m_{\rm geo} = \prod_i m_i^{p_i}\, ,
\end{equation}
\noindent  where $m_{\rm geo}$ is the geometric mean of stellar masses in an embedded cluster and $m_i$ are the possible fragment masses and $p_i$ their relative weights with $\sum_i p_i = 1$. 
It represents the characteristic scale around which the stellar mass distribution is centered. Here, $m_i$ is the sequence of stellar masses defined by Eq.~\ref{eq:m_i}, and $p_i$ is the corresponding normalized distribution defined by Eq.~\ref{eq:p_i}. Taking the natural logarithm of Eq.~\ref{eq:geo_mass} transforms the geometric sum into an arithmetic one \citep{Siegel1942}:
\begin{equation}\label{eq:frag_constraint}
    \ln m_{\rm geo} = \ln\left(\prod_i m_i^{p_i}\right) = \sum_i p_i \ln m_i = \langle \ln m\rangle\, ,
\end{equation}
where the quantity inside the angle brackets denotes the expectation value of the natural logarithm of the stellar mass. We therefore see that the natural logarithm of the geometric mean is equal to the expectation value of the logarithm of the stellar masses.
In the case of a continuous distribution, the geometric mass can be expressed as:
\begin{equation}\label{eq:geomass_cont}
    m_{\rm geo} = \exp\left( \frac{1}{N} \int_{m_{\rm min}}^{m_{\rm max}} \, \xi(m) \, \ln m \, \diff m \right) \, .
\end{equation}
The expression follows from substituting Eq.~\ref{eq:p_i} to the right-hand side of Eq.~\ref{eq:geo_mass}, replacing the discrete sum $\sum_i$ with the continuous integral $\int_{m_{\rm min}}^{m_{\rm max}} \diff m$, and the discrete variable, $m_i$, with the continuous variable, $m$.

\subsection{The sIMF from the Maximum Entropy Principle}\label{sec:sIMF_MaxEnt}

Assuming that star formation is primarily dominated by gravitational fragmentation (Sec.~\ref{sec:fragmentation}), let
us consider the probability distribution that maximizes the entropy, $S$
(Eq.~\ref{eq:entropy}). Along with the distribution constraint from
Eq.~\ref{eq:probability_axiom}, the Lagrangian of the star-forming system should
be subject to the constraint from Eq.~\ref{eq:frag_constraint},
i.e.\footnote{Note that Eq.~\ref{eq:logmass_minimization}, when expressed as a
function of $\xi(m_i)$, becomes:
\begin{equation}
    \sum_i \frac{\xi(m_i) \, \Delta m_i}{N}\ln m_i = N \langle \ln m\rangle \, .
\end{equation}}:
\begin{align}
\sum_i p_i - 1 &= 0 \label{eq:probability_condition}  \\
\sum_i p_i \ln m_i -  \langle \ln m\rangle   &= 0 \, . \label{eq:logmass_minimization}
\end{align}

Using Eqs.~\ref{eq:probability_axiom}~and~\ref{eq:logmass_minimization} as
constraints associated to the Lagrange multipliers $\lambda_0,\lambda_1$ we can
define the Lagrangian governing the sIMF:

\begin{align}
    \mathcal{L} = &-\sum_i p_i \ln p_i - (\lambda_0 - 1)\left(\sum_i p_i - 1\right) \nonumber\\ 
     &- \lambda_1\left(\sum_i p_i \ln m_i - \langle \ln m\rangle\right) \, ,\label{eq:IMFlagrangian}
\end{align}
and then taking the derivative with respect to each $p_i$:
\begin{equation}\label{eq:Lminimization}
    \frac{\partial \mathcal{L}}{\partial p_i} = -(\ln p_i + 1) - (\lambda_0 - 1) - \lambda_1 \ln m_i = 0 \, ,
\end{equation}
simplifies to:
\begin{equation}
    \ln p_i = -\lambda_0 - \lambda_1 \ln m_i \, ,
\end{equation}
and now the probability distribution can be solved as:
\begin{equation} \label{eq:pi_distribution}
    p_i = e^{-\lambda_0}\, e^{-\lambda_1 \ln m_i} = e^{-\lambda_0} m_i^{-\lambda_1} \propto
    m_i^{-\alpha} \, .
\end{equation}

Recall from Eq.~\ref{eq:canonicalIMF} that $\xi(m)$ is generally represented as
a power law. We may consider a simpler form of the sIMF, for example, the
original single power law \citep[][S55]{Salpeter1955} derived for the mass range
$0.4<m/{\rm M}_{\odot}<10$:
\begin{equation} \label{eq:SalpeterIMF}
    \xi_{\rm S55}(m) = k_{\star} m^{-\alpha} \, ,
\end{equation}
where $k_{\star}$ is the normalization constant which ensures
Eq.~\ref{eq:IMFdistribution} and Eq.~\ref{eq:opt_mass} are fulfilled, i.e., the
integral of the sIMF over the full mass range returns the total number of stars,
while the integral of the mass-weighted sIMF, with integrand $m \, \xi(m)$,
returns the embedded cluster stellar mass, $M_{\rm ecl}$.
Comparing Eq.~\ref{eq:SalpeterIMF} with Eq.~\ref{eq:pi_distribution}, we notice that the first Lagrange multiplier, $\lambda_1$, coincides with the power-law slope of the sIMF, $\alpha$. The normalization constant of the sIMF, $k_{\star}$, is related, together with Eq.~\ref{eq:p_i}, to the zeroth Lagrange multiplier, via:

\begin{equation}\label{eq:kstar}
    k_{\star} = N e^{- \lambda_0} \, .
\end{equation}

The Lagrange multipliers encapsulate the physics of the system, and can be determined empirically or theoretically based on the physical properties of the system. 

Additional physics or constraints (e.g., metallicity, cosmic~rays, or additional phenomena, see Sec.~\ref{sec:fragmentation}) may be included as additional Lagrange multipliers. In our case, we consider only
one constraint (i.e., $r=1$ from Eqs.~\ref{eq:lagrangian} and
\ref{eq:Lm_constraints}). Our specific single constraint is $M_{\rm ecl}$. We discuss this next.

\subsection{Optimal Sampling as a Consequence of Maximum Entropy}\label{sec:OptSamp}

In the previous section, by minimizing a Lagrangian constrained by fragmentation (Eq.~\ref{eq:IMFlagrangian}) we obtained the shape of the sIMF normalized to unity, i.e. the $p_i$ distribution from Eq.~\ref{eq:pi_distribution}.
Without additional constraints in the Lagrangian, the distribution $p_i$ must also satisfy the condition that the total mass it generates coincides with $M_{\rm ecl}$. That is, the full distribution must be consistent with the total stellar mass of the embedded cluster:

\begin{align}
    \sum_i m_i\,\xi(m_i) \, \Delta m_i & = M_{\rm ecl} \, , \label{eq:discrete_opt_mass}\\
    \sum_i m_i\, p_i  & = \frac{M_{\rm ecl}}{N} \, , \label{eq:totmass}
\end{align}
where Eq.~\ref{eq:totmass} follows from substituting Eq.~\ref{eq:p_i} into Eq.~\ref{eq:discrete_opt_mass}. This is one of the optimal sampling conditions.

If we discretize the single power-law prescription for the sIMF (Eqs.~\ref{eq:SalpeterIMF} and \ref{eq:kstar} with Eq.~\ref{eq:xi_i}) we find that:
\begin{equation}\label{eq:discrete_singleplaw}
    \xi_i = \xi(m_i) = \frac{N}{\Delta m_i} p_i
= \frac{N e^{-\lambda_0}}{\Delta m_i} m_i^{-\lambda_1}\,.
\end{equation}
Given that $p_i$ must be normalized so that $\sum_{i=1}^{\rm N} p_i = 1$, we find:
\begin{equation}
    e^{-\lambda_0} \sum_{i=1}^{\rm N} m_i^{-\lambda_1} = 1 \, ,
\end{equation}
or:
\begin{align}
    e^{-\lambda_0} &= \frac{1}{\sum_{i=1}^{\rm N} m_i^{-\lambda_1}} \, ,\\
    \lambda_0 &= \ln \left(\sum_{i=1}^{\rm N} m_i^{-\lambda_1}\right) \, . \label{eq:constraint}
\end{align}
We see that $\lambda_0$ and $\lambda_1$ are not independent, and that $\lambda_0$ is defined  by $\lambda_1$ along with the exact sequence of masses in the distribution.
Let us introduce:
\begin{equation}
      Z_1 = \sum_{i=1}^{\rm N} m_i^{-\lambda_1} = e^{\lambda_0} \, ,
\end{equation}
meaning that the distribution $p_i$ can be rewritten as:
\begin{equation}\label{eq:pi_solution}
    p_i = \frac{1}{Z_1} m_{i}^{-\lambda_1}\, .
\end{equation}
The first Lagrange multiplier condition (Eq.~\ref{eq:logmass_minimization}) is equivalent to:
\begin{equation} \label{eq:first_lagrange_multiplier_condition}
    \langle \ln m \rangle = \frac{1}{Z_1}\sum_{i=1}^{\rm N} m_i^{-\lambda_1} \ln m_i \, .
\end{equation}

Let us add one additional star to the ensemble, characterized by mass $m_{\rm N+1}$ and distribution $p_{\rm N+1}$. 

In order to satisfy the Maximum Entropy principle, this new ensemble must also satisfy Eq.~\ref{eq:IMFlagrangian}. The minimization of the Lagrangian (Eq.~\ref{eq:Lminimization}) requires that the new distribution satisfies Eq.~\ref{eq:first_lagrange_multiplier_condition}: 
\begin{equation} \label{eq:new_first_lagrange_multiplier_condition}
    \langle \ln m \rangle' = \frac{1}{Z_1'}\sum_{\rm i'=1}^{\rm N+1} m_{\rm i'}^{-\lambda_1} \, \ln m_{\rm i'} \, ,
\end{equation}
where $\langle \ln m \rangle '$ represents the new characteristic mass, the index $\rm i' = \{\rm 1, ..., N+1\} = \{\rm i, N+1\}$ spans one extra element, and:
\begin{equation}
    e^{\lambda_0'} = Z_1' = \sum_{\rm i'=1}^{\rm N+1} m^{-\lambda_1}_{\rm i'} = Z_1 + m_{\rm N+1}^{-\lambda_1} = e^{\lambda_0} + m_{\rm N+1}^{-\lambda_1} \, .
\end{equation}
The new number of stars, $N'$, and new embedded cluster mass, $M_{\rm ecl}'$, will be:
\begin{align}
    N' &= N + 1 \, , \label{eq:Nplus1}\\
    M_{\rm ecl}'&= M_{\rm ecl} + m_{\rm N+1} \, , \label{eq:Mecl_plus1} 
\end{align}
which then implies:
\begin{align}
    N' &= (N+1)\sum_{\rm i'=1}^{\rm N+1} p_{\rm i'} \, , \\
    M_{\rm ecl}'&=  (N+1) \sum_{i'=1}^{\rm N+1} m_{\rm i'} \, p_{\rm i'} \, ,
\end{align}
where $p_{\rm i'}$ is re-normalized so that the sum of $i$ from 1 to $\rm N+1$ adds to 1:
\begin{equation}
    p_{\rm i'}' = \frac{1}{Z_1'} \, m_{\rm i'}^{-\lambda_1} \, .
\end{equation}

Let us further simplify Eq.~\ref{eq:new_first_lagrange_multiplier_condition}:
\begin{equation}
    \langle \ln m \rangle ' = \frac{1}{Z_1 + m_{\rm N+1}^{-\lambda_1}}\left(m_{\rm N+1}^{-\lambda_1} \, \ln m_{\rm N+1} + Z_1 \,  
    \langle \ln m \rangle \right) \, ,
\end{equation}
which can be regrouped to:
\begin{equation} \label{eq:our_important_finding}
    Z_1\left(\langle \ln m\rangle' - \langle \ln m\rangle  \right) = m_{\rm N+1}^{-\lambda_1} \left( \ln m_{\rm N+1} - \langle \ln m \rangle ' \right) \, .
\end{equation}
This expression defines a necessary condition to ensure that the new star belongs to the same ensemble obtained from Eq.~\ref{eq:IMFlagrangian}, without violating the maximum entropy principle. 
Eq.~\ref{eq:our_important_finding} is equivalent to asserting that only stars drawn from the deterministic optimal sampling sequence satisfy the entropy-maximizing framework under fixed constraints. For any new massive star, $m_{\rm N+1}$, appended to the distribution $\{ m_i\}$, the following two inequalities hold $\left(\langle \ln m\rangle' > \langle \ln m\rangle  \right)$ and $\left( \ln m_{\rm N+1} > \langle \ln m \rangle ' \right)$. 
The term $\left( \ln m_{\rm N+1} - \langle \ln m \rangle ' \right)$ identifies to what extent the newly added mass deviates from the new characteristic mass, the geometric mean mass. The term $\left(\langle \ln m\rangle' - \langle \ln m\rangle  \right)$ identifies the deviation of the new characteristic mass from the previous one. Under the first Lagrange multiplier condition (i.e., the slope of the sIMF) the ratio between the two differences  is identified uniquely by the ratio $\frac{Z_1}{m_{\rm N+1}^{-\lambda_1}}$, where the numerator is the normalization of the original sIMF and the denominator is the newly added most massive star. 

If both the initial distribution, $p_{\rm i}$, and the augmented distribution, $p_{\rm i}'$, abide by the maximum entropy principle, then all stars in both distributions must be described by the same power law slope, our first Lagrange multiplier $\lambda_1$. 
Given that the sum of both  
\begin{equation}
    \sum_{\rm i=1}^{\rm N} p_{\rm i}= \frac{1}{Z_1}\sum_{\rm i=1}^{\rm N} m_{\rm i}^{-\lambda_1}  =1
\end{equation}
and 
\begin{equation}
    \sum _{\rm i'=1}^{\rm N+1} p_{\rm i '} = \frac{1}{Z_1'} \left(\sum _{\rm i=1}^{\rm N} m_{\rm i}^{-\lambda_1} + m_{\rm N+1}^{-\lambda_1}\right) = 1
\end{equation} 
must total to unity, the respective normalization constants $Z_1$ and $Z_1'$ are also bound exactly through the newly added star $m_{\rm N+1}$. 
These constraints are equivalent to the optimal sampling constraint from Sec.~\ref{sec:SF}, where we found that solving the integrals of the optimal sampling conditions simultaneously returns both the most massive star, $m_{\rm max}$, and the sIMF normalization constant, $k_{\star}$.

\begin{figure*}[t]
    \centering
    \begin{subfigure}[t]{\columnwidth}
        \centering
        \includegraphics[width=\linewidth]{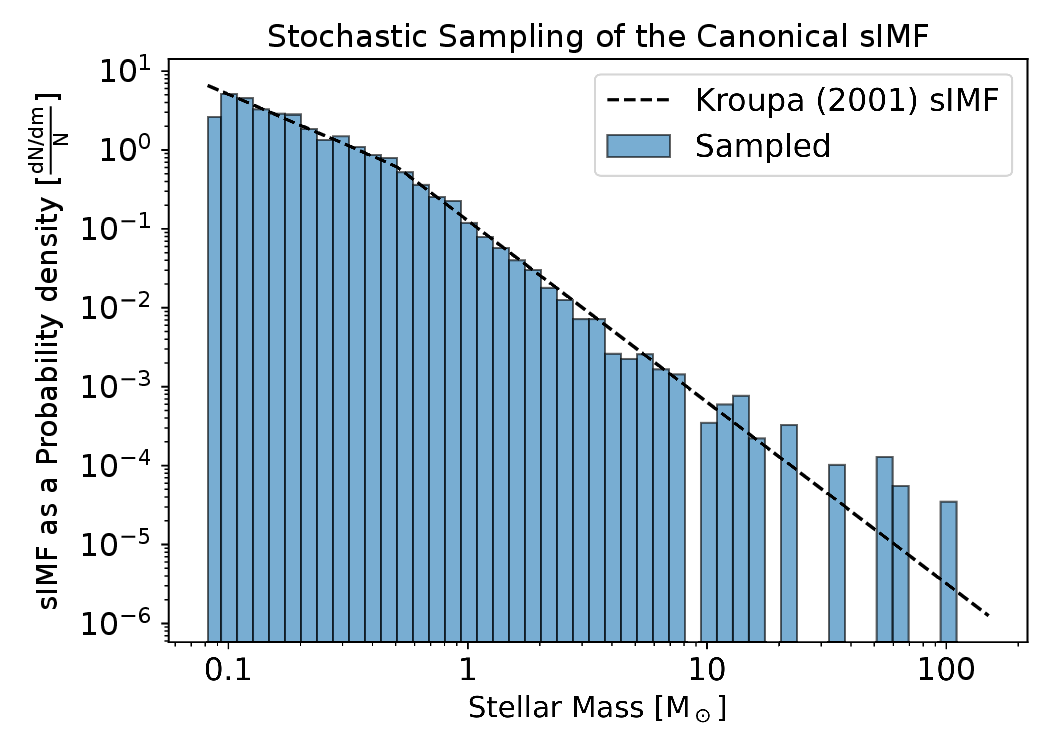}
        \caption{One random sampling realization of the canonical sIMF whose integral is normalized to 1. Logarithmic $\Delta m$ spacing divided into 50 bins from 0.08 to 150 solar masses. For this realization, the embedded cluster mass is $M_{\rm ecl} = 1083 \, {\rm M}_{\odot}$, close within Poisson noise to the $M_{\rm ecl}$ from Fig.~\ref{fig:optimal}.}
        \label{fig:random}
    \end{subfigure}
    \hfill
    \begin{subfigure}[t]{\columnwidth}
        \centering
        \includegraphics[width=\linewidth]{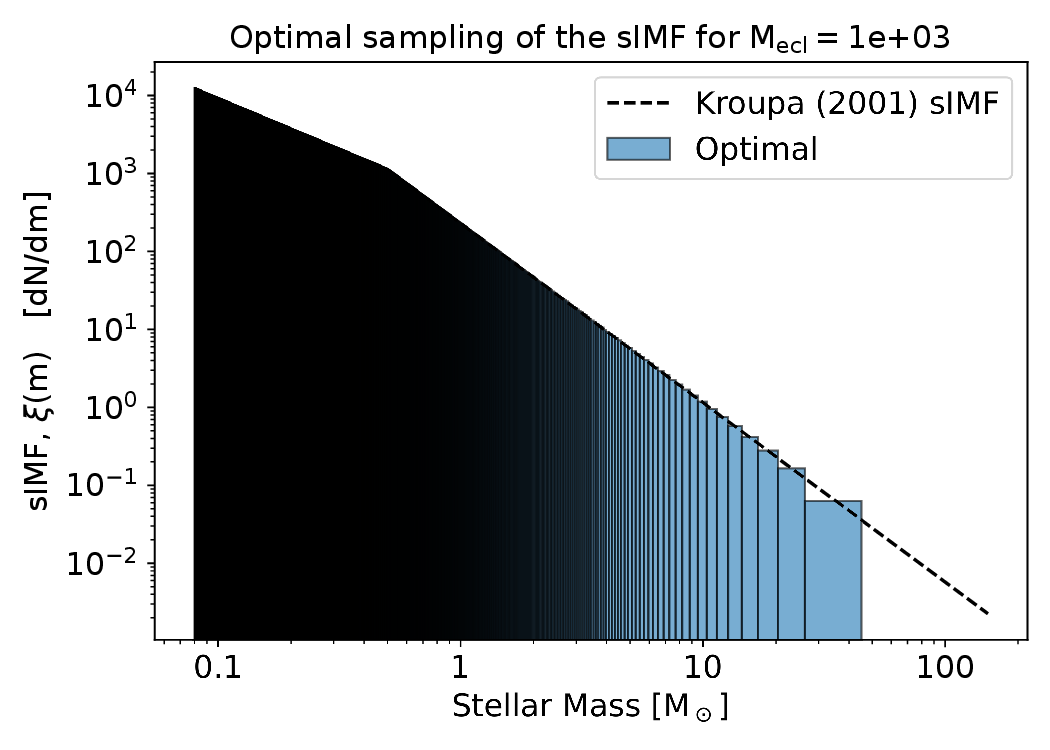}
        \caption{Optimal sampling of a canonical sIMF for an embedded cluster of total stellar mass $M_{\rm ecl} = 10^3 \, {\rm M}_{\odot}$. Each bin contains a single star. The most massive star obtained via optimal sampling contains 45~${\rm M}_{\odot}$. Note that in the stochastic sampling (Fig.~\ref{fig:random}), stars above 45~${\rm M}_{\odot}$ may appear sparsely. }
        \label{fig:optimal}
    \end{subfigure}
\end{figure*}

Eq.~\ref{eq:our_important_finding} expresses a necessary condition for the augmented ensemble (our thought experiment where we add an hypothetical new most massive star). This necessary condition ensures the solution, which provides the distribution of the system, keeps maximizing the entropy under the same set of constraints. For fixed $\lambda_1$ and fixed fragmentation constraint $\langle \ln m\rangle$, there is at most one value of $m_{\rm N+1}$ that satisfies this condition. In other words, once the slope of the sIMF and the characteristic mass set by fragmentation are fixed, the mass of the next star in the sequence is determined by the requirement that the updated ensemble continues to solve the variational problem expressed in Eq.~\ref{eq:IMFlagrangian}.

The only constraint we introduced is the normalization by the total embedded cluster mass, $M_{\rm ecl}$, formed by an ensemble. This is the first optimal sampling condition.
At the discrete level, this is expressed by Eq.~\ref{eq:discrete_opt_mass}, which is equivalent to Eq.~\ref{eq:totmass} once we rewrite it in terms of the normalized distribution $p_i$.

The constraint from Eq.~\ref{eq:totmass} therefore fixes the mean stellar mass of the ensemble. 
The probability normalization (Eq.~\ref{eq:probability_axiom}), the fragmentation constraint (Eq.~\ref{eq:frag_constraint}), and the total mass condition (Eq.~\ref{eq:totmass}), taken together, identify a unique discrete sequence $\{m_i\}$ and corresponding probabilities $\{p_i\}$. 
This is implicitly the second optimal sampling condition.

Among all discrete distributions $\{p_i\}$ that satisfy these constraints, the MaxEnt solution, Eq.~\ref{eq:pi_solution}, 
has the largest entropy $S = -\sum_i p_i \ln p_i$. Any attempt to modify one of the masses $m_i$ while keeping the constraints fixed produces a more structured distribution with a lower value of $S$. In this sense, the optimal sampling configuration is the least biased realization compatible with the fragmentation physics and the prescribed $M_{\rm ecl}$, because it reproduces the required statistics of the sIMF and introduces no additional substructure.

Suppose now that the added star $m_{\rm N+1}$ does not follow the discrete sIMF, $\xi_i$, as given by Eq.~\ref{eq:discrete_singleplaw}. Then $m_{\rm N+1}$ does not satisfy Eq.~\ref{eq:our_important_finding}, and there is no choice of $Z_1'$ for which the augmented ensemble solves the same variational problem with the same $\lambda_1$ and fragmentation constraint. The updated distribution would fail to minimize the entropy functional under the original constraints of Eq.~\ref{eq:IMFlagrangian}. To recover an entropy–maximizing configuration one would need to change the constraints (for example a different $\lambda_1$ or a different sequence $\{m_i\}$), in which case $m_{\rm N+1}$ would not belong to the original embedded cluster defined by the initial fragmentation physics.

Given a MaxEnt distribution $p(m)$, the associated deterministic sequence of stellar masses is obtained via the cumulative distribution function (CDF),
\begin{equation}
    F(m) \equiv \int_{m_{\rm min}}^{m} p(m')\,\diff m' \, .
    \label{eq:CDF}
\end{equation}
On the mass range $[m_{\rm min},m_{\rm max}]$, $F(m)$ is strictly increasing and continuous, so it admits an inverse, the quantile function $Q(u) \equiv F^{-1}(u)$ for $0 < u < 1$. 
The choice of $u \in (0,1)$ is due to the fact that the CDF satisfies $F(m_{\rm min})=0$ and $F(m_{\rm max})=1$ for a normalized distribution, so its inverse, $Q(u)=F^{-1}(u)$, is defined on the unit interval.
For a population of $N$ stars we define the ordered masses through the quantile map,
\begin{equation}
    m_i \equiv Q\left(\frac{i}{N}\right) \, ,
    \label{eq:quantile_sequence}
\end{equation}
where $i=1,\dots,N $, as before.
By construction, each interval $[m_{i-1},m_i]$ contains the probability from Eq.~\ref{eq:p_i}:
\begin{equation}
    \int_{m_{i-1}}^{m_i} p(m)\,\diff m
    = F(m_i) - F(m_{i-1})
    = \frac{i}{N} - \frac{i-1}{N}
    = \frac{1}{N} \, ,
\end{equation}
with $m_0 \equiv m_{\rm min}$. Multiplying by $N$ and using Eq.~\ref{eq:p_i}, we obtain:
\begin{equation}
    \int_{m_{i-1}}^{m_i} \xi(m)\,\diff m
    = N \int_{m_{i-1}}^{m_i} p(m)\,\diff m
    = 1 \, ,
    \label{eq:opt_sampling_second_condition}
\end{equation}
which is exactly the optimal sampling condition: each interval in the ordered mass sequence contains one star. The most massive star is $m_{\rm max} = m_{\rm N} = Q(1)$ and occupies the last bin alone.

\begin{figure*}
    \centering
\includegraphics[width=0.49\linewidth]{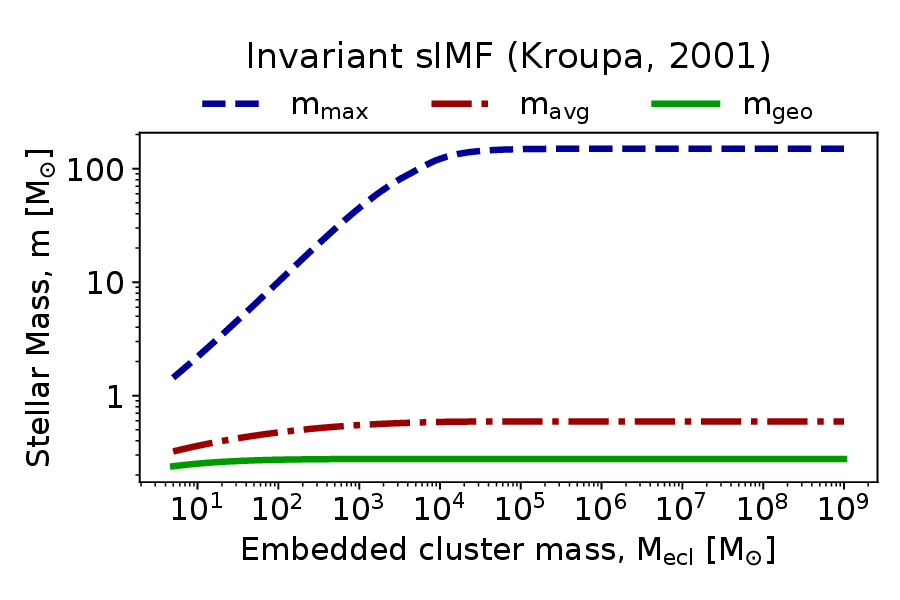}
\includegraphics[width=0.49\linewidth]{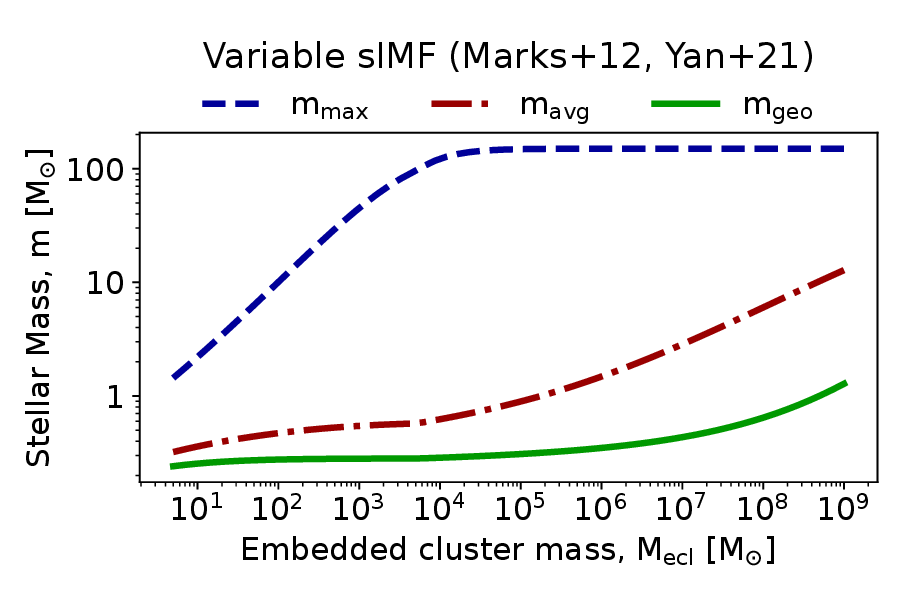}~

    \caption{A comparison between 3 relevant masses as a function of embedded cluster mass, $M_{\rm ecl}$: the most massive star constrained by optimal sampling ($m_{\rm max}$, \emph{blue dashed line}), the average stellar mass in the embedded cluster ($m_{\rm avg}$, \emph{red dot-dashed line}) and the geometric mean mass ($m_{\rm geo}$, \emph{solid green line}) computed according to Eq.~\ref{eq:geomass_cont}. The blue curve is the $m_{\rm max}-M_{\rm ecl}$ relation. Up until $M_{\rm ecl}\lesssim 10^4 \, M_{\odot}$ where $m_{\rm max}\lesssim 150 \, M_{\odot}$, the average and geometric masses increase. On the left panel, for $M_{\rm ecl}\gtrsim 10^4 \, M_{\odot}$ the average and geometric masses remain constant. On the right panel, they increase because of the variability of the sIMF. We assumed the specific variability as given in \citet{marks+12, yan+21}, and computed with \pyIGIMF $\,$ \citep{Gjergo+2025} for solar metallicity. In this case, the sIMF starts to become top-heavy and consequently, both average and geometric mass increase.}
    \label{fig:geomeanmax}
\end{figure*}

Note that Eq.~\ref{eq:opt_sampling_second_condition} holds for any choice of sIMF, not only for the fragmentation scenario we derived in Sec.~\ref{sec:fragmentation}. The optimal sampling framework can therefore be applied to broken power–law sIMFs or other functional forms once a target sIMF has been specified.

Equation~\ref{eq:opt_sampling_second_condition} shows that the sequence $\{m_i\}$ constructed from the MaxEnt density through Eq.~\ref{eq:quantile_sequence} reproduces the optimal sampling prescription. Because $F$ is strictly monotonic, the quantile map is invertible, and the ordered sequence $\{m_i\}$ is unique for a given $p(m)$ and $\rm N$. Any attempt to shift one of the masses $m_i$ while keeping the same $p(m)$ and the same number of stars would break the equal–probability condition and therefore the one–star–per–interval condition in Eq.~\ref{eq:opt_sampling_second_condition}.

The conclusion is that, once the fragmentation constraint and the total stellar mass $M_{\rm ecl}$ are fixed, the MaxEnt formalism selects a unique optimal sampling sequence. The last bin contains a single most massive star, and all other stars occupy the remaining bins in a way that is fully determined by the sIMF and the constraints, without stochastic deviations from the optimal sequence.

Hence, as also shown most recently in \citet{yan+23}, and supported by several works summarized in \citet{kroupa+2024}, 
we have found that the only configuration consistent with the entropy-maximizing variational principle is one in which stellar masses follow the sequence prescribed by the sIMF (or equivalently, by $p_i$), without any stochastic deviations. 
The sIMF is therefore not a probabilistic distribution from which stellar masses may be drawn, but a descriptive representation of how the physics of the molecular clump  populates the permitted configuration of the system.

\section{Visualizations} \label{sec:visualizations}

To illustrate the difference between stochastic and optimal sampling, we show a
realization of stochastic sampling from the sIMF for an embedded cluster with a
mass of $M_{\rm ecl} \approx 10^3\,{\rm M}_{\odot}$ (Fig.~\ref{fig:random}), and
the corresponding sIMF discretized via optimal sampling for a similarly massive
cluster with average solar metallicity (Fig.~\ref{fig:optimal}). 
In both figures, instead of showing a single power law, we decided to apply optimal sampling to the most common sIMF distribution, the canonical IMF \citep{Kroupa2001}, which is often taken to be universal for main-sequence galaxies. 
It can be seen from Fig.~\ref{fig:random} that stochastic sampling may exhibit gaps in the stellar mass range, and the sIMF
value in each bin is subject to Poisson noise. In contrast, optimal sampling from Fig.~\ref{fig:optimal}
reproduces the sIMF as closely as possible, without noise or discontinuities. 

 It is important to recall that the distributions in Figs.~\ref{fig:random}~and~\ref{fig:optimal} captures the full ensemble of stars formed by a molecular clump over its lifetime of about 1 Myr. It cannot be fully observed while star-formation is ongoing, and it cannot be observed soon after when star-formation ends, because the resulting embedded cluster undergoes rapid dynamical evolution \citep[see][]{dinnbier+2022}.

In Fig.~\ref{fig:geomeanmax}, we compare the newly proposed
fragmentation scale mass (Eq. \ref{eq:geomass_cont}, here referred to as the geometric mean mass, $m_{\rm geo}$), to the mass of the most massive star in the embedded cluster, $m_{\rm max}$, and to the average stellar mass in the embedded cluster $m_{\rm avg}$. 
We consider two scenarios, one adopting an invariant canonical sIMF \citep{Kroupa2001} and the other adopting the variable sIMF scheme developed in \citet{marks+12, yan+21}, and most recently summarized in \citet{Gjergo+2025}.

In the invariant sIMF case, both the geometric mean mass, $m_{\rm geo}$, and the arithmetic mean mass, $m_{\rm avg}$, show a mild increase with $M_{\rm ecl}$ until $m_{\rm max}$ reaches its theoretical upper limit. Once the full sIMF range is populated, these two averages approach a constant value, as expected from Eq.~\ref{eq:geomass_cont}. 
The most massive star increases with $M_{\rm ecl}$ until it saturates at $150\,{\rm M}_{\odot}$. However, massive stars are rare and therefore contribute little to the typical stellar mass captured by the geometric mass. In Eq.~\ref{eq:geo_mass}, they enter only through $\ln m$, and the probability weight at the high-mass end is low. As a result, $m_{\rm geo}$ depends weakly on $M_{\rm ecl}$. Once $m_{\rm max}$ reaches its physical ceiling, the upper cutoff stops changing. The relative weights across the sIMF then remain fixed, so $\langle\ln m\rangle$ and therefore $m_{\rm geo}$ plateau.

In the variable sIMF case, which becomes top-heavy for $M_{\rm ecl} \gtrsim 10^4\,{\rm M}_{\odot}$, both $m_{\rm max}$ and $m_{\rm geo}$ increase accordingly.
This result appears consistent with thermal Jeans mass fragmentation observed across clumps of varying mass \citep[e.g.,][]{Zhang2009, Morii2024, Li2024}. All quantities are computed assuming solar metallicity with the public software, \pyIGIMF\ (\citealt{Gjergo+2025} \footnote{\url{https://github.com/egjergo/pyIGIMF}}). As explained in Sec.~\ref{sec:intro}, the theoretical upper limit for the most massive stellar mass is assumed to be $m_{\rm max *} \approx 150 \, M_{\odot}$. The asymptotic behavior of the true most massive star, $m_{\rm max}$, identifies when this upper limit has been reached. This blue curve is also known as the $m_{\rm max}-M_{\rm ecl}$ relation \citep[for a review, see ][]{kroupa+2024}.

\section{Comparing Optimal Sampling with Observations and Alternative Interpretations}
\label{sec:discussion}

Stars condense out of gas in molecular clumps. It is natural to hypothesize that the sIMF originates in the fluid dynamics of these environments. Undeniably, the interstellar medium behaves as a turbulent fluid. It is therefore often assumed that the shape of the sIMF comes from the kinetic energy cascade across spatial scales \citep[e.g.,][]{Padoan2002}. The corresponding power spectrum of turbulence follows:
\begin{equation}
    P(k) \propto k^{-\gamma_{\rm p}} \, ,
\end{equation}
with $\gamma_{\rm p} = 5/3$ for subsonic turbulence and $\gamma_{\rm p} = 2$ for supersonic turbulence. Subsonic turbulence produces a power law, while supersonic turbulence produces log-normal distributions \citep[for a review, see][]{Bastian2010}.

Under the premise that turbulence directly defines the shape of the sIMF, said sIMF is often interpreted as a probability density function with a log-normal and a power-law segment, being the premise of gravo-turbulent theory (e.g., \citealt{Padoan2002, HennebelleChabrier2011}), which, however, observations and simulations disfavor, as shown next.
These perspectives led to gravo-turbulent theory, which rests on the assumption that turbulence ultimately gives rise to the sIMF. This leads to a stochastic interpretation of the sIMF. Optimal sampling is in direct antithesis to this scenario.

According to the gravo-turbulent theory of star formation, density peaks
arise in regions compressed by shocks, and subsequently collapse
into protostars if they are sufficiently dense \citep{Padoan2002,Elmegreen2011}. The
IMF is then expected to emerge from the statistical distribution of these peaks as a log-normal function for low-mass stars and into the sub-stellar regime. However, simulations suggest that low-density peaks are frequently destroyed by subsequent shocks before collapse can occur, casting doubt on the ability of turbulence to
regulate the sIMF \citep{BertelliMotta+2016}. With high resolution MHD
simulations, \cite{Guszejnov2022} shows that the initial turbulence level has little impact on the shape of the sIMF, which also challenges the 
gravo-turbulent interpretation.

When stars form, the gas has already settled into a very cold ($< 20 \, {\rm K}$) low-energy state \citep{Goodman+1993} and is embedded in highly coherent filamentary structures \citep{Andre+2014}. These patterns contradict the expectations of turbulent fragmentation. Recent high-resolution observations show that stars with $m<1\,{\rm M}_{\odot}$ tend to form in narrow ($\approx 1\,$pc-wide) filaments with near-regular spacing between protostars and strong coherence in both spatial and velocity space \citep[e.g.,][]{Andre+2014, hacar+2017}. 

Such structures are irreconcilable with a stochastic interpretation that stars arise from density peaks in a turbulent medium. 
First, filaments condense. Such filaments are characterized by local density fluctuations \citep{Andre+2019}. Gas in the filaments falls toward the local overdensities, in which protostars form. Each embedded cluster emerging from this protostar formation progressively breaks apart the tenuous infalling filaments in a continuous process until the accumulating stellar feedback halts star formation. 

Each protostar is thereby launched onto a ballistic orbit within the evolving gravitational potential. This implies close multi-proto-stellar encounters in the dense innermost region of the evolving embedded cluster, leading to significant three-body-encounter driven ejections of the most massive objects. Thus, very young clusters eject a significant fraction of their most massive stars \citep[e.g.,][]{ohkroupa2012, ohkroupa2016}. This has been confirmed using Gaia data on the star-burst cluster R136 \citep{GT2021, Sana+2022, Stoop+2024}.

Massive stars tend to form from central hubs where multiple filaments converge \citep[e.g.,][]{YangLiuTej+2023}. In such hubs, the fragmentation of dense cores is accompanied by a highly dynamic environment and evolution dominated by strong accretion flows \citep[e.g.,][]{Liu2012a,Liu2012b,hacar+2025}. Under these conditions, if protostars are treated as point-like objects in stellar dynamical terms, dense cores may coalesce before protostars
condense within the clumps \citep[e.g.][]{dib+2007}. The collapse of a stellar core occurs within a timescale of about $10^5\,$yr, see \citealt{kroupa+2024}.
Otherwise, massive stellar cores reach their final mass after prolonged accretion \citep[e.g.,][]{Sollins2005,GM2023}. As a result, the mass function of the cores, and consequently of the protostars and stars, becomes top-heavy in the densest hubs, i.e. the central regions of dense embedded clusters \citep[see][and references therein]{kroupa+2024}.

Sub-stellar objects—brown dwarfs and some very low-mass stars with $m < 0.2\,{\rm M}_{\odot}$ also form a continuous mass distribution that appears to be a power-law with index $\alpha_0\approx 0.3$ \citep[][and references therein]{kroupa+2013, kroupa+2024}
but must originate (owing to their very different initial binary properties)  from a distinct formation channel stemming from the fragmentation of perturbed outer accretion disks termed ``secondary fragmentation'' \citep{thies+2010, thies+2015}. Noteworthy in this context is that the two branches (direct fragmentation of the clump into stars yielding the sIMF) and the secondary fragmentation overlap such that in the mass range $0.06-0.2\,{\rm M}_{\odot}$ the two sIMFs overlap.  

The above discussion supports the notion that different physical constraints and thus different Lagrange multipliers affect the accessible sIMF states such that the power-law index ($\lambda_1$ in Eq.~\ref{eq:pi_distribution}) differs over these mass ranges. 

In statistical physics, the Maximum Entropy principle is equivalent to the condition of thermodynamic equilibrium. For an isolated system at fixed total energy, entropy is maximized at equilibrium \citep{landau1980}. In systems coupled to a heat bath, equilibrium corresponds to the minimization of free energy, or equivalently, to the maximization of the total entropy of the system and reservoir. Physically, this principle captures the natural tendency of systems to evolve toward thermodynamically-stable macro-states. These are not necessarily states of minimum energy. Rather, they are the states accessible by the largest number of micro-states, as quantified by the system's sIMF. We stress once again that by micro-state (see Sec.~\ref{sec:SF}) we refer to properties of individual stars. Specifically, their mass distribution.

The equilibrium of such a thermodynamic system closely mirrors the behavior of optimal sampling: the formation of a massive star is only permitted once the lower-mass portion of the sIMF, relative to that star, has been fully populated.  
The physical interpretation of this is that as the clump begins to collapse under self gravity, it begins forming low-mass stars in the filaments within it, which, as the collapse proceeds into denser conditions and more gas falls in along forming new filaments, proceeds to form the more massive and most massive stars in the core of the hub. These ultimately halt further gas inflow through their ionizing radiation and winds unless they are ejected from the forming embedded cluster as appears to have occurred two times in the Orion Nebula Cluster \citep{kroupa+2018a}.

\section{Conclusions}
\label{sec:conclusion}

In this work we address why the sIMF should be interpreted as an optimal distribution function rather than as a probability density function. This distinction has important consequences for the evolution of stellar systems and for the interpretation of extragalactic observations \citep[e.g.][]{haslbauer+2024}.

We have shown that, for a fragmenting molecular clump, the least biased (minimum information, maximum-entropy) mass distribution under the fragmentation constraint is a power-law sIMF\footnote{
That means, under optimal sampling, we only need the total mass of the stellar population, $M_{\rm ecl}$, to describe the full distribution. In contrast, stochastic sampling requires the knowledge of every stellar mass to fully describe any single realization of a stellar population.}. We then proved that, once probability normalization (Eq.~\ref{eq:probability_condition}) and the total stellar mass, $M_{\rm ecl}$ (e.g., Eq.~\ref{eq:opt_mass}), are imposed, the maximum-entropy solution selects a unique ordered stellar mass sequence that coincides with the optimal sampling construction, which requires the stellar mass distribution to be ordered (Eq.~\ref{eq:opt_sampling_second_condition}, first proposed in \citealt{kroupa+2013}). Optimal sampling thus emerges as the natural deterministic realization of the sIMF, because it does not introduce bias or intrinsic scatter in the sIMF , and agrees better with observations compared to stochastic sampling \citep{yan+23, Gjergo+2025, kroupa+2024}.

This is not surprising, because the Maximum Entropy principle is formulated as a variational problem, just like other foundational theories in physics. In this context, \citet{CarcassiAidala2020} showed that classical mechanics is associated with high-entropy systems, whereas quantum mechanics corresponds to low-entropy systems.

In particular, the mass distribution of newly formed stars has a natural analogue in the Boltzmann distribution: individual degrees of freedom follow deterministic laws, but the ensemble displays maximal uncertainty about the state of any one of them. In the same way, the sIMF in our framework is not a sampling probability distribution but a deterministic realization of the physical state of the system, selected by macroscopic (whole–cluster) constraints while individual stellar states need not be known.

We would like to clarify that the application of the MaxEnt Principle in this work does not imply that the star-forming system is in thermodynamic equilibrium. Rather,  MaxEnt is a principle of inference, and applies to information entropy. It identifies the most probable distribution given a set of macroscopic constraints. It is a variational principle applied to probability distributions, not to physical states. The term ``equilibrium'', therefore, refers to the statistical equilibrium of the inferred distribution under the fragmentation constraint, not to a relaxation of the dynamical system.

We obtained the result that the sIMF is an optimal distribution function
by applying the Maximum Entropy principle on a scenario where stars form from filaments hierarchically. Maximum entropy selects the state of maximal uncertainty (i.e., minimal information content) compatible with the known constraints, and therefore yields the least biased distribution. These constraints can be refined in future work to include additional physics (for example further fragmentation, metallicity, turbulence, or cosmic rays), but the dominant process remains dynamical, accretion–driven fragmentation \citep[see][]{kroupa+2024}.

The optimal sampling framework was originally identified by \citet{kroupa+2013}. With optimal sampling, the sIMF is populated without gaps, so each molecular clump produces a stellar population that follows the shape of the sIMF, just as gas in thermal equilibrium follows the Boltzmann distribution. In both cases, the equilibrium distribution is reproducible and fixed by macroscopic constraints, despite the complexity of the underlying dynamics. Only a fraction of the clump gas mass participates in star formation, and this available mass, the embedded cluster mass, sets the normalization of the sIMF. If the embedded cluster mass is small,  the available gravitational binding energy and associated accretion rates are insufficient to populate the high-mass end of the sIMF. In this regime, the sIMF is not stochastically sampled. Consequently, massive stars do not form, not due to chance, but because the system lacks the mass and corresponding gravitational energy required to populate that part of the distribution.

The empirically established broken power-law form of the sIMF (Eq.~\ref{eq:canonicalIMF}) may reflect the action of different fragmentation channels over distinct mass ranges. As proposed by \citet{kroupa+2024}, the flatter low-mass slope is attributed to filamentary fragmentation, which depends sensitively on metallicity. Stars more massive than $1\,{\rm M}_{\odot}$ form predominantly in dense hubs where filaments intersect, where accretion-driven fragmentation in these high-density regions becomes the dominant process. 
There may be exceptions to these patterns, such as the Central Molecular Zone \citep{zhang+2025} and the Galactic nuclear star cluster \citep{bartko+2010}. However, the majority of stars form in the classic conditions outlined in \citet{kroupa+2024}.

We conclude that optimal sampling is the natural consequence of treating the sIMF as the final ``equilibrium'' state of a deterministic fragmentation process, in which the physical state of the molecular clump fixes the ordered sequence of stellar masses in the embedded cluster. This result is obtained under the assumption that the observer has maximal uncertainty (or equivalently, minimum information) about the internal state of the system, i.e., by applying the Maximum Entropy principle. In this view, the sIMF plays a role analogous to an equilibrium distribution in thermodynamics: complex out-of-equilibrium dynamics converge to a reproducible configuration determined by a small set of macroscopic constraints.

In practice, we implemented our analysis by extremizing the entropy functional, $\mathcal{L}$ (Eq.~\ref{eq:IMFlagrangian}), with calculus of variations. In our framework, the non-trivial physical constraint is cloud fragmentation. The additional constraints we considered are sIMF normalization (Eq.~\ref{eq:probability_axiom}, also Eq.~\ref{eq:probability_condition}) and the total stellar mass condition (Eq.~\ref{eq:logmass_minimization}). These two conditions together supply the only necessary constraints that result in the sIMF being optimally sampled (Eq.~\ref{eq:opt_sampling_second_condition}). Independent of the detailed fragmentation driver, observations and theoretical models consistently obtain a power-law sIMF for $m > 0.5 \, {\rm M}_{\odot}$. Under these constraints the Maximum Entropy solution selects a unique ordered mass sequence, which coincides with the optimal sampling configuration and contains no stochastic deviations from the sIMF.

In this analysis we focused on the low embedded cluster mass regime. We highlight a non-trivial coincidence, also noted by \citet{Gjergo+2025}, that once $m_{\rm max}$ reaches its physical ceiling, the sIMF becomes top-heavy at higher $M_{\rm ecl}$. The slope in this top-heavy regime has been constrained empirically using independent data. This behaviour suggests that, once a star-forming overdensity provides a sufficiently large mass reservoir, the high-mass end is populated more efficiently, shifting the sIMF towards a top-heavy form.

\begin{acknowledgements}
We thank the anonymous referee and Alessandro Bressan for their thoughtful and constructive feedback, which helped improve the quality and clarity of this manuscript. E.G. and Z.Z. acknowledge the support of the National Natural Science Foundation of China (NSFC) under grants NOs. 1251101411, 12173016, 12041305.
E.G. and Z.Z. acknowledge the Program for Innovative Talents, Entrepreneur in Jiangsu. 
P.K. acknowledges support through the DAAD-Eastern-European Exchange programme between Bonn and Prague, and through the grant No. 26-217745 from the Czech Grant Agency.
\end{acknowledgements}

\appendix
\section{Error order of discretizing the sIMF}

Note that the continuous and discrete forms of the sIMF integrals, i.e.:
\begin{equation}
    \int \xi(m) \, \diff m \, \approx \, \sum_i \xi_i \Delta m_i
\end{equation}
has a small Big-O error\footnote{``Big-O'' is a way to describe how an error term evolves with some small parameter. A small Big-O error means that the neglected term goes to zero quickly while decreasing the parameter.} because the step $\Delta m_i$ is adaptive. This produces a non-uniform quadrature that samples finely for large $\xi(m)$, and in turn improves the convergence. Similarly to adaptive quadrature, this approximation has an efficient convergence improving as
\begin{equation}
    \mathcal{O}\left(\frac{1}{N^2}\right) \, .
\end{equation}

\bibliography{ms2025-0298main}{}
\bibliographystyle{raa}

\label{lastpage}

\end{document}